\documentclass[aps,prl,preprint,onecolumn,final,11pt,groupedaddress,showpacs]{revtex4-1}
\newcommand{\bi}{\bibitem}
\usepackage[latin1]{inputenc}
\usepackage[english]{babel}
\usepackage{graphicx}
\usepackage{psfrag}
\usepackage{amssymb}
\usepackage{amsmath}
\usepackage{amscd}
\usepackage{eucal}
\usepackage{color}
\usepackage{bm}
\newcommand{\Kf}{K\hspace{-0.8mm}f}
\input{epsf}
\begin{document}
\renewcommand{\figurename}{Figure}

\title{The Key Player Problem in Complex Oscillator Networks and Electric Power Grids:  Resistance Centralities Identify Local Vulnerabilities}

\author{M.~Tyloo}
\affiliation{Institute of Physics, EPF Lausanne, CH-1015 Lausanne, Switzerland }
\affiliation{School of Engineering, University of Applied Sciences of Western Switzerland HES-SO CH-1951 Sion, Switzerland}

\author{L.~Pagnier}
\affiliation{Institute of Physics, EPF Lausanne, CH-1015 Lausanne, Switzerland }
\affiliation{School of Engineering, University of Applied Sciences of Western Switzerland HES-SO CH-1951 Sion, Switzerland}

\author{Ph.~Jacquod}
\affiliation{School of Engineering, University of Applied Sciences of Western Switzerland HES-SO CH-1951 Sion, Switzerland}

\date{\today}

\begin{abstract}
Identifying key players in a set of coupled individual systems is a fundamental problem in network theory~\cite{Bal06,Bor06,Fli13}. 
Its origin can be traced back to social sciences and the problem led to  
ranking algorithms based on graph theoretic centralities~\cite{Bol14}.
Coupled dynamical systems differ from social 
networks in that, first, they are characterized by degrees of freedom with a deterministic dynamics and second
the coupling between individual systems is a well-defined function of those degrees of freedom. 
One therefore expects the 
resulting coupled dynamics, and not only the network topology, to also determine the key players. 
Here, we investigate synchronizable network-coupled dynamical systems such 
as high voltage electric power grids and coupled oscillators on complex networks. 
We search for network nodes which, once perturbed by a local noisy disturbance, generate the 
%%%% which are most vulnerable to....
largest overall transient excursion away from synchrony. A spectral decomposition of the network coupling matrix leads to an 
elegant, concise, yet accurate solution to this identification problem.
We show that, when the internodal coupling matrix is Laplacian, these key players are peripheral in the 
sense of a centrality measure defined from effective resistance distances.
For linearly coupled dynamical systems such as weakly loaded electric power grids or consensus algorithms, the nodal 
ranking is efficiently obtained through a single Laplacian
matrix inversion, regardless of the operational synchronous state. We call the resulting ranking index {\it LRank}.
For heavily loaded electric power grids or coupled oscillators systems closer to the transition to synchrony, 
nonlinearities render the nodal ranking 
dependent on the operational synchronous state. In this case a weighted Laplacian matrix inversion gives another
ranking index, which we call {\it WLRank}. Quite surprisingly, we find that LRank provides a faithful ranking even 
for well developed coupling nonlinearities, corresponding to oscillator angle differences up to $\Delta \theta \lesssim 40^o$
approximately. 
\end{abstract}

\maketitle

\section{Introduction}

Because of growing electric power demand, increasing difficulties with building new lines and the emergence of 
intermittent new renewable energy sources, electric power systems are more often operated closer to their 
maximal capacity~\cite{Kra16,NAS16}. Accordingly, their operating state, its robustness against potential disturbances and
its local vulnerabilities need to be assessed more frequently and precisely. Furthermore, because electricity markets become more 
and more integrated, it is necessary to perform these assessments over geographically larger areas.
Grid reliability is commonly assessed against
$n-1$ feasibility, transient stability and voltage stability, by which one means that 
a grid is considered reliable if (i) it still has an acceptable operating state
after any one of its $n$ components fails, (ii) that acceptable state is reached from the original state 
following the transient dynamics generated by the component failure and (iii) the new operating state is robust against 
further changes in operating conditions such as changes in power productions and loads. This $n-1$
contingency assessment is much harder to implement in real-time for a power grid loaded close to its capacity where
the differential equations governing its dynamics become nonlinear -- the fast, standardly used linear approximation 
breaks down as the grid is more and more heavily loaded. 
Nonlinear assessment algorithms have significantly longer runtimes,
which makes them of little use for short-time evaluations.
In worst cases, they sometimes even do not converge. In short,
heavily loaded grids need more frequent, more precise reliability 
assessments which are however harder to obtain, precisely because the loads are closer to the grid capacities. 

Developing real-time procedures for $n-1$ contingency assessment  requires new, innovative algorithms. One appealling avenue
is to optimize contingency ranking~\cite{Fli13} to
try and identify a subset of $n_s<n$ grid components containing all the potentially 
critical components. The $n-1$ contingency assessment 
may then focus on that subset only, with a significant gain in runtime if $n_s \ll n$. Identifying such a subset requires 
a ranking algorithm for grid components, following 
some well-chosen criterion. Procedures of this kind have been 
developed in network models for social and computer sciences, biology and other fields, in the context of the 
historical and fundamental problem of identifying the {\it key players}~\cite{Bal06,Bor06,Sol01,Mon06}.  
They may be for instance the players who, once removed, 
lead to the biggest changes in the other player's activity in game theory, or to the biggest structural change 
in a social network. That problem has been addressed with the introduction of graph theoretic centrality 
measures~\cite{Bol14,Bon87} which order nodes from the most "central" to the most "peripheral" -- 
in a sense that they themselves define. A plethora of centrality indices have been introduced and discussed in the literature on network theory~\cite{Bol14,Bon87}, culminating with 
PageRank~\cite{Bri98}. The latter ranks nodes in a network according to the 
stationary probability distribution of a Markov chain on the network,
accordingly it gives a 
meaningful ranking of websites under the reasonable assumption that websurfing is a random process.
Their computational efficiency makes PageRank, as well as other purely graph theoretic indicators
very attractive to identify key players on complex networks. 
It is thus quite tempting to apply purely graph theoretic methods to identify fast and reliably key players in network-coupled
dynamical systems.

%%% remove lines from here
Processes such as web crawling for information retrieval are essentially random diffusive walks on a complex network,
with no physical conservation law beyond the conservation of probability. The situation is similar for 
disease~\cite{Kit10} or rumor~\cite{Bor12} spreading, and 
for community formation~\cite{Gir02}
where graph theoretic concepts of index, centrality, betweenness, coreness and so forth have been successfully applied
to identify tightly-bound communities.
%%% to here ? (if shorter version needed)
Coupled dynamical systems  such as complex supply networks~\cite{Her06}, electric power grids~\cite{Bia08}, consensus algorithm networks~\cite{Lyn97} or more generally network-coupled oscillators~\cite{Kur75,Ace05}
are however fundamentally different. There,  
the randomness of motion on the network giving e.g. the Markovian chain at the core
of PageRank is replaced by a deterministic dynamics supplemented by physical conservation laws that cannot be neglected.
Pure or partially extended graph theoretic methods have been applied in
vulnerability investigations of electric power grids~\cite{Bom13,San13,Has17}, and investigations of cascades of failures
in coupled communication and electric power networks~\cite{Bul10,Bas13}. They have however
been partially or totally invalidated by investigations on
more precise models of electric power transmission that take fundamental physical laws into
account (in this case, Ohm's and Kirchhoff's laws)~\cite{Hin10,Kor17}.
It is therefore doubtful that purely topological graph theoretic descriptors are able to identify the potentially 
critical components in deterministic, network-coupled dynamical systems. Purely graph-theoretic approaches need to be extended
to account for physical laws~\cite{Bom13}.

\begin{figure}[t]
 \begin{center}
\includegraphics[width=450px]{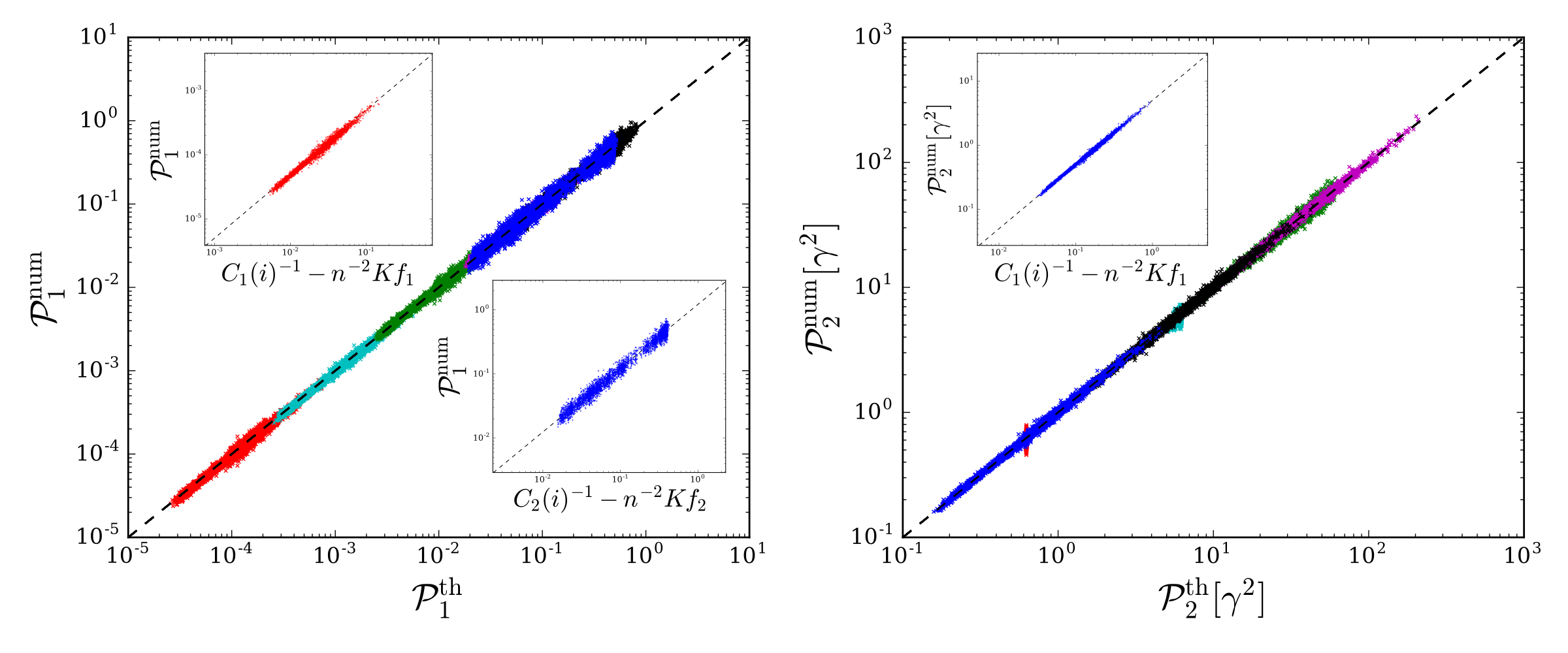}
  \caption{\footnotesize Comparison between theoretical predictions and numerical results for both performance measures
  ${\cal P}_1$ and ${\cal P}_2$ defined in Eqs.~(\ref{eq:performance}). 
  Each point corresponds to a noisy disturbance 
  on a single node of the European electric power grid sketched in Fig.~\ref{fig5}a (supplementary materials, materials and methods)  and governed by Eq.~(\ref{eq:swing}). The time-dependent disturbance $\delta P_i(t)$ is defined by an Ornstein-Uhlenbeck 
  noise of magnitude $\delta P_0=1$ and correlation 
  time $\gamma \tau_0=4\cdot 10^{-5}$ (red crosses), $4\cdot 10^{-4}$ (cyan), $4\cdot 10^{-3}$ (green), $4\cdot 10^{-2}$ (purple),  
  $4 \cdot 10^{-1}$ (black) and $4$ (blue). Time scales are defined by the ratio of damping to inertia coefficients
  $\gamma=d_i/m_i=0.4 s^{-1}$ which is assumed constant with $d_i=0.02s$.
  The insets show ${\cal P}_1$ and ${\cal P}_2$ as a function of the resistance distance-based 
  graph theoretic predictions of Eqs.~(\ref{eq:p1p2}) valid in both limits of very large and very short noise correlation time
  $\tau_0$. Not shown is the limit of short $\tau_0$ for ${\cal P}_2$, which gives a node-independent
  result, Eq.~(\ref{eq:p2}).} %Error bars correspond to one standard deviation calculated from 40 realizations of noise.
%  Where not shown, they are smaller than the size of the markers.}
  \label{fig1}
 \end{center}
\end{figure}

Here, we give an elegant solution to the key player problem for a family of deterministic,
network-coupled dynamical systems related to the Kuramoto model~\cite{Kur75,Ace05}. While we focus mostly on
high voltage electric power grids whose swing dynamics, under the lossless line approximation, 
is given by a second-order version of the Kuramoto model~\cite{Bia08,Dor13a}, we show that our approach also applies to 
other, generic models of network-coupled oscillators. 
Key players in such systems can be defined in various ways. For instance, they can be  
identified by an optimal geographical distribution of system parameters
such as inertia, damping or natural frequencies or alternatively as those 
whose removal leads to the biggest change in operating state. In this article we define the key players 
as those nodes where a local disturbance leads to the largest 
network response. There are different measures to quantify the magnitude of the transient response, such as
nadir and maximal rate of change of the network-averaged frequency~\cite{Pag17,Guo18} or other
dynamical quantities such as  network susceptibilities~\cite{Man17} and the wave dynamics following 
disturbances~\cite{Tam18}. Here, we quantify the total transient excursion through 
performance measures that are time-integrated quadratic forms in the system's degrees of freedom (supplementary materials, materials and methods). Anticipating on results 
to come, 
Fig.~\ref{fig1} illustrates the excellent 
agreement between analytical theory and numerical calculations for such performance measures. 
Particularly interesting is that 
in both asymptotic limits of quickly and slowly decorrelating noisy disturbance, 
the performance measures are simply expressed in terms of the {\it resistance centrality}~\cite{Ste89,Bra05}, which is
a variation of the closeness centrality~\cite{Bol14} based on resistance distances~\cite{Kle93}. 
This is shown in the insets of Fig.~\ref{fig1}. Our main finding is that the resistance centrality is the 
relevant quantity to construct ranking algorithms in network-coupled dynamical systems.

%%%%%%%%
%%%%%%%%

\section{The Model and Approach}
We consider network-coupled dynamical systems defined by sets of differential equations of the form
\begin{eqnarray}\label{eq:swing}
m_i \ddot{\theta}_i + d_i \dot{\theta}_i = P_i - \sum_j b_{ij}\sin(\theta_i-\theta_j) \, ,\;\, i= 1,...,n .
\end{eqnarray} 
The coupled individual systems are
oscillators with a compact, angle degree of freedom $\theta_i \in (-\pi,\pi]$. Their
uncoupled dynamics are determined 
by natural frequencies $P_i$, inertia coefficients $m_i$ and damping coefficients $d_i$. Because the degrees of freedom are
compact, the coupling between oscillators needs to be a periodic function of angle differences and here we keep only
its first Fourier term. 
The coupling between pairs of oscillators is defined on a network whose 
Laplacian matrix has elements $\mathbb{L}_{ij}^{(0)} =-b_{ij} $ if $i\neq j$ and $\mathbb{L}_{ii}^{(0)}=\sum_{k \ne i} 
b_{ik}$. Without inertia, $m_i = 0$ $\forall i$, Eq.~(\ref{eq:swing}) gives the celebrated Kuramoto 
model on a network with edge weights $b_{ij}>0$, $\forall i,j$~\cite{Kur75,Ace05}. With inertia on certain nodes, it is an approximate model for the swing dynamics of 
high-voltage electric power grids in the lossless line limit~\cite{Bia08,Ber81,Dor13a}.
When angle differences are small, 
a linear approximation $\sin(\theta_i-\theta_j) \simeq \theta_i-\theta_j$ is justified, giving first- (without) or second-order (with inertia)
consensus dynamics~\cite{Lyn97}.

When the natural frequencies $P_i$ are not too large, synchronous solutions exist that satisfy 
Eq.~(\ref{eq:swing}) with $ \ddot{\theta}_i =0$ and $\dot{\theta}_i=\omega_0$, $\forall i$. Without loss of generality,
one may consider Eq.~(\ref{eq:swing}) in a frame rotating with the angular frequency $\omega_0$
in which case such synchronous states correspond to 
stable fixed points with $\dot{\theta}_i =0$. We consider a fixed point with angle coordinates 
$\bm \theta^{(0)} = (\theta_1^{(0)}, \ldots , \theta_n^{(0)})$ corresponding to
natural frequencies $\bm P^{(0)} = (P_1^{(0)}, \ldots , P_n^{(0)})$, to which we add a time-dependent disturbance, $P_i (t) =  P_i^{(0)}+ \delta P_i(t)$.
Linearizing the dynamics about that solution, Eq.~(\ref{eq:swing}) becomes
\begin{eqnarray}\label{eq:swing_lin}
m_i \delta \ddot{\theta}_i + d_i \delta \dot{\theta}_i = \delta P_i(t) - \sum_j b_{ij}\cos(\theta_i^{(0)}-\theta_j^{(0)}) 
(\delta \theta_i - \delta \theta_j) \, ,\;\, i= 1,...,n ,
\end{eqnarray} 
where $\delta \theta_i(t)=\theta_i(t)-\theta_i^{(0)}$.
This set of coupled differential equations governs the 
small-signal response of the system corresponding to weak disturbances. The couplings are defined 
by a weighted Laplacian matrix $\mathbb{L}_{ij}(\bm \theta^{(0)})=-b_{ij}\cos(\theta^{(0)}_i-\theta^{(0)}_j)$ if $i\neq j$ and $\mathbb{L}_{ii}(\bm \theta^{(0)})=\sum_k b_{ik}\cos(\theta^{(0)}_i-\theta^{(0)}_k)$
which contains information on
both the topology of the network and the operational state of the system. This weighted Laplacian matrix significantly differs from 
the network Laplacian $\mathbb{L}^{(0)}$
when angle differences between coupled nodes are large. 

We assess the nodal vulnerability of the system defined in Eq.~(\ref{eq:swing}) via the magnitude of the transient dynamics 
determined by Eq.~(\ref{eq:swing_lin}) under a time-dependent disturbance $\delta P_i(t)$.
We take the latter as an Ornstein-Uhlenbeck noise on the natural frequency of a single node,
with vanishing average, $\overline{\delta P_i(t)} = 0$, variance $\delta P_0^2$ and correlation time $\tau_0$, 
$\overline{\delta P_i(t_1)\delta P_j(t_2)}=\delta_{ik} \, \delta_{jk} \, \delta P_0^2\exp[-|t_1-t_2 |/\tau_0]$.
It is sequentially applied on each of the 
$k=1, \ldots n$ nodes.  This noisy test disturbance is designed to investigate network properties on different time scales
by varying $\tau_0$ and identify the set of
most vulnerable nodes as those where the system's response to $\delta P_k(t)$ is largest.
We quantify the magnitude of the response to the disturbance with the following two performance measures~\cite{Tyl18}
\begin{subequations}\label{eq:performance}
\begin{equation}\label{eq:p1perf}
{\mathcal{P}_1}=\lim_{T\rightarrow\infty}T^{-1}\sum_i\int_0^T |\delta \theta_i(t) - \Delta(t) |^2 {\rm d}t \; ,
\end{equation}
\begin{equation}
{\mathcal{P}_2}=\lim_{T\rightarrow\infty}T^{-1}\sum_i\int_0^T |\delta \dot{\theta}_i(t) - \dot{\Delta}(t) |^2 {\rm d}t \; .
\end{equation}
\end{subequations} 
They are similar to performance measures based on ${\cal L}_2$-norms previously considered in the context of electric power
networks~\cite{Teg15,Sia14,Sia16,Poo17,Pag17,Col18}
but differ from them in two respects.  First, here we 
 subtract the averages $\Delta (t) = n^{-1}
\sum_j \delta \theta_j(t)$ and $\dot\Delta (t)  = n^{-1} \sum_j \delta \dot\theta_j(t)$ because the synchronous state does not change under a constant angle shift. Without that subtraction, artificially large performance measures may be obtained, which reflect 
a constant angle drift of the synchronous operational state
and not a large transient excursion. Second, we divide ${\mathcal{P}_{1,2}}$ by $T$
before taking $T \rightarrow \infty$ because we consider a noisy disturbance that is not limited in time and which would otherwise
lead to diverging values of ${\mathcal{P}_{1,2}}$.

\section{Performance measures and Resistance Centralities}

The performance measures 
${\mathcal{P}_{1,2}}$ can be computed analytically from Eq.~(\ref{eq:swing_lin}) via Laplace transforms
(supplementary materials, materials and methods). For uniform damping and inertia, i.e. $d_i=d=\gamma m_i$, $\forall i$, in the two limits of long and short noise correlation time $\tau_0$, they can be expressed in terms of the resistance centrality of the node
$k$ on which the noisy disturbance acts and of graph topological indices called generalized Kirchhoff indices~\cite{Kle93,Tyl18}.
Both quantities are based on the resistance distance, which gives the effective 
resistance $\Omega_{ij}$ between any two nodes $i$ and 
$j$ on a fictitious electrical network where each edge is a resistor of magnitude given by the inverse edge 
weight in the network defined by the 
weighted Laplacian matrix.  One obtains
\begin{equation}\label{eq:rdistance}
\Omega_{ij}(\bm \theta^{(0)}) =\mathbb{L}_{ii}^\dagger(\bm \theta^{(0)}) +
\mathbb{L}_{jj}^\dagger(\bm \theta^{(0)})-\mathbb{L}_{ij}^\dagger(\bm \theta^{(0)})-\mathbb{L}_{ji}^\dagger(\bm \theta^{(0)}) \, ,
\end{equation}
where $\mathbb{L}^\dagger$ denotes the Moore-Penrose pseudo-inverse of $\mathbb{L}$~\cite{Kle93}.
The resistance centrality of the $k^{\rm th}$ node is then defined as $C_1(k) = [n^{-1} \sum_j \Omega_{jk}]^{-1}$. It measures how
central node $k$ is in the electrical network, in terms of its average resistance distance to all other nodes. A
network descriptor, the Kirchhoff index is further defined as~\cite{Kle93} 
\begin{equation}\label{eq:kf1}
\Kf_1 \equiv \sum_{i<j} \Omega_{ij} \, .
\end{equation}

Generalized Kirchhoff indices $\Kf_p$ and resistance centralities $C_p(k)$ can be defined analogously 
from the $p^{\rm th}$ power of the weighted Laplacian matrix, which is also a Laplacian matrix 
(supplementary materials, materials and methods).
In terms of these quantities, the performance measures defined in Eqs.~(\ref{eq:performance}) 
depend on the value of the noise correlation time 
$\tau_0$ relative to the different time scales in the system. The latter are the ratios
$d/\lambda_\alpha$ of the damping coefficient $d$ with the nonzero eigenvalues $\lambda_\alpha$, $\alpha=2, \ldots n$, of 
$\mathbb{L}(\bm \theta^{(0)})$ and the inverse ratio $\gamma^{-1}=m/d$ of damping to inertia coefficients. The performance measures take in particular the asymptotic values 
\begin{subequations}\label{eq:p1p2}
\begin{equation}\label{eq:p1}
\mathcal{P}_1 = \left\{
  \begin{array}{lr}
    \big(\delta P_0^2\tau_0 \big/ d ) \left({C_1^{-1}(k)}-n^{-2}\Kf_1\right) \; \; ,\;  \tau_0\ll d/\lambda_\alpha, \gamma^{-1} \\
    \delta P_0^2\left( {C_2^{-1}(k)}-n^{-2}\Kf_2\right) \; \; ,\;  \tau_0\gg d/\lambda_\alpha, \gamma^{-1} 
  \end{array}
\right.
\end{equation}
\begin{equation}
\mathcal{P}_2 = \left\{
  \begin{array}{lr}
    \big(\delta P_0^2\tau_0 \big/ dm \big) \big( n-1 \big) \big/ n  \; \;  ,\;  \tau_0\ll d/\lambda_\alpha, \gamma^{-1}  \\
    \big(\delta P_0^2 \big/ d\tau_0 \big) \left( {C_1^{-1}(k)}-n^{-2}\Kf_1\right) \; \; ,\;  \tau_0\gg d/\lambda_\alpha, \gamma^{-1}  ,\label{eq:p2}
  \end{array}
\right.
\end{equation}
\end{subequations}
in the two limits when $\tau_0$ is the smallest or the largest time scale in the system. After averaging over the location $k$ of
the disturbed node, $\overline{C_{1,2}} = 2 \Kf_{1,2}/n^2$, and one recovers the results of Refs.~\cite{Sia14,Sia16,Tyl18} for the 
global robustness of the system.

These results are remarkable : they show that the magnitude of the transient excursion 
under a local noisy disturbance is given by either of the generalized resistance centralities $C_{1}(k)$ or 
$C_{2}(k)$ of the perturbed node and 
the generalized Kirchhoff indices $\Kf_{1,2}$. The latter are global network descriptors and are therefore fixed 
in a given network with fixed operational state. One concludes that perturbing the less central nodes -- those with largest inverse
centralities $C_{1,2}^{-1}(k)$ -- generates the largest transient excursion. The asymptotic  
analytical results of Eqs.~(\ref{eq:p1p2}) are corroborated by numerical
results in the insets of Fig.\ref{fig1}, obtained directly from Eq.~(\ref{eq:swing}), i.e. without the linearization
of Eq.~(\ref{eq:swing_lin}). The validity of the general analytical 
expressions valid for any $\tau_0$ (supplementary materials, materials and methods) is further confirmed in the main
panel of Fig.~\ref{fig1}, and by further numerical results obtained for different networks shown in the supplementary materials, materials and methods.

\begin{figure}[t]
 \begin{center}
\includegraphics[width=160px]{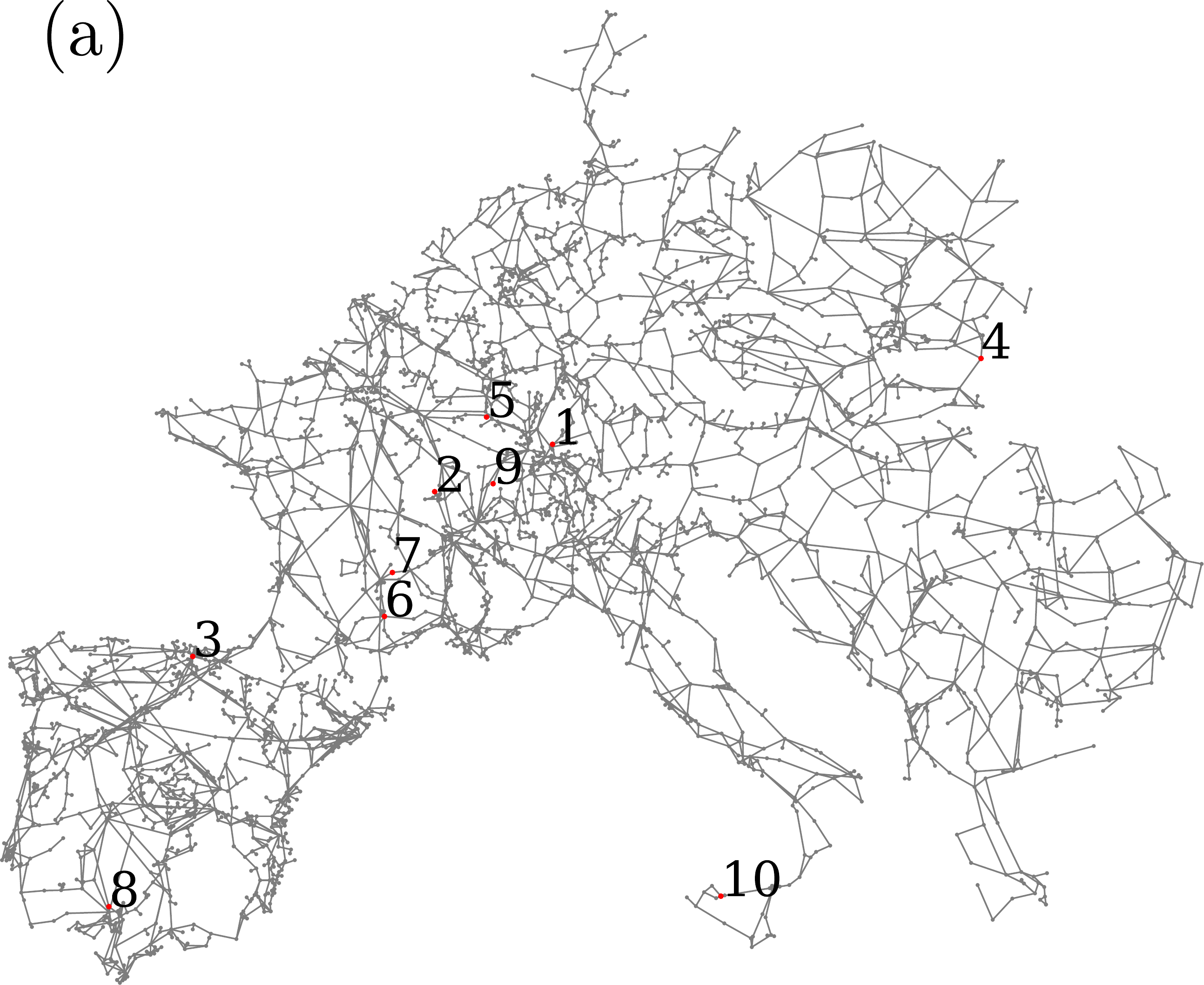}
\includegraphics[width=290px]{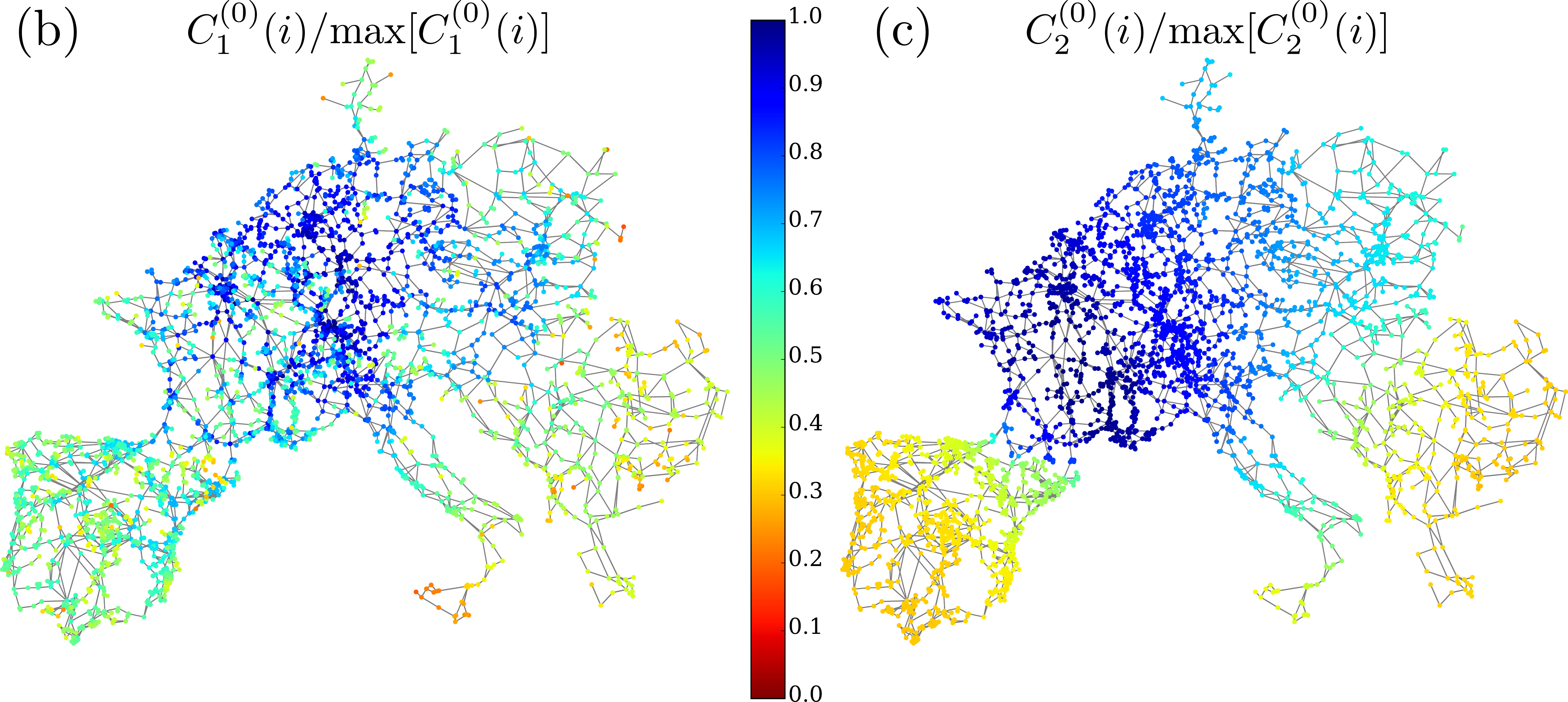}
  \caption{\footnotesize (a) Topology of the European electric power grid (supplementary materials, materials and methods)
  and location of the ten test nodes listed in Table~\ref{table1}. 
  Normalized generalized resistance centralities $C_1^{(0)}(i)$ (b), and $C_2^{(0)}(i)$ (c) 
  for the network Laplacian matrix of the European electric power grid.}
  \label{fig5}
 \end{center}
\end{figure}

The generalized resistance centralities and Kirchhoff indices appearing in Eqs.~(\ref{eq:p1p2}) depend on the operational state via
the weighted Laplacian $\mathbb{L}(\bm \theta^{(0)})$. For a narrow distribution of natural frequencies $P_i \ll \sum_j b_{ij}$, 
$\forall i$,
angle differences between coupled nodes remain small, and the weighted Laplacian is close to the network 
Laplacian, $\mathbb{L}(\bm \theta^{(0)}) \simeq \mathbb{L}^{(0)}$. The resistance centralities $C_{1}^{(0)}$ and $C_{2}^{(0)}$ for the  network 
Laplacian of the European electric power grid are shown in Fig.~\ref{fig5}. For both centralities, 
the less central nodes are dominantly located in the Balkans and Spain. Additionally, for $C_1^{(0)}$, nodes in Denmark and
Sicily are also among the most 
peripheral. The general pattern of these most peripheral nodes looks very similar to the pattern of most
sensitive nodes numerically found in Ref.~\cite{Gam17}, 
and includes in particular many, but not all dead ends, which have been numerically found
to undermine grid stability~\cite{Men14}. 

The asymptotic results of Eqs.~(\ref{eq:p1p2}), together with the numerical results of Fig.~\ref{fig1} make a strong point that 
nodal sensitivity to fast or slowly decorrelating noise disturbances can be 
predicted by generalized resistance centralities. One may wonder 
at this point how generalized resistance centralities differ in that prediction from other, more common centralities such as geodesic
centrality, nodal degree or PageRank. Table~\ref{table1} compares these centralities to each other and to the performance 
measures corresponding to slowly decorrelating noisy disturbances acting
on the ten nodes shown in Fig.~\ref{fig5}a. As expected from Eq.~(\ref{eq:p1p2}),
$\mathcal{P}_{1}$ and $\mathcal{P}_{2}$ are almost perfectly correlated
with the inverse resistance centralities $C_2^{-1}$ and $C_1^{-1}$ respectively, but with no other centrality metrics.
For the full set of nodes of the Europen electric power grid, we found 
Pearson correlation coefficients $\rho(\mathcal{P}_{1},C_2^{-1})=0.997$, and $\rho(\mathcal{P}_{2},C_1^{-1})=0.975$
fully corroborating the prediction of Eq.~(\ref{eq:p1p2}).
\begin{table}[t]
\begin{tabular}{|c|c|c|c|c|c|c|c|c|c|c|c|c}
\hline
node \# & ${C_\mathrm{geo}}$ & Degree & PageRank & ${C_1}$ & ${C_2}$  &   $\mathcal{P}^{\mathrm{num}}_1$ & $\mathcal{P}^{\mathrm{num}}_2$  $[\gamma^2]$  \\ \hline
1&7.84&4&3024&31.86&5.18&0.047&0.035\\ \hline
2&6.8&1&2716&22.45&5.68&0.021&0.118\\ \hline
3&5.56&10&896&22.45&2.33&0.32&0.116\\ \hline
4&4.79&3&1597&21.74&3.79&0.126&0.127\\ \hline
5&7.08&1&1462&21.74&5.34&0.026&0.125\\ \hline
6&4.38&6&2945&21.69&5.65&0.023&0.129\\ \hline
7&5.11&2&16&19.4&5.89&0.016&0.164\\ \hline
8&4.15&6&756&19.38&1.83&0.453&0.172\\ \hline
9&5.06&1&1715&10.2&5.2&0.047&0.449\\ \hline
10&2.72&4&167&7.49&2.17&0.335&0.64 \\\hline

\end{tabular}
\caption{\footnotesize Centrality metrics and performance measures $\mathcal{P}_{1,2}$ for the European electric power grid 
(supplementary materials, materials and methods) with noisy disturbances with large correlation time $\tau_0$ 
applied on the nodes shown in Fig.~\ref{fig5}a. The performance measures $\mathcal{P}_{1}$ and $\mathcal{P}_{2}$ are 
almost perfectly correlated
with the resistance centralities $C_2$ and $C_1$, 
but neither with the geodesic centrality, nor the degree, nor  PageRank.}
\label{table1}
\end{table}

\section{Ranking of local vulnerabilities}

Once a one-to-one relation between the generalized resistance centralities $C_{1}(k)$ and $C_{2}(k)$ of the disturbed node $k$
and the
magnitude of the induced transient response is established, ranking of nodes from most to least critical 
is tantamount to ranking them from smallest to largest $C_1$ or $C_2$. From Eqs.~(\ref{eq:p1p2}), 
which of these two centralities is relevant depends on whether one is interested 
(i) in the transient response under fast or slowly decorrelating noise,
or (ii) in investigating transient behaviors for angles (using the performance measure ${\cal P}_1$) or frequencies (${\cal P}_2$).
Quite interestingly, while this gives a priori four different rankings, 
Eqs.~(\ref{eq:p1p2}) lead to only two rankings, either based
on $C_1^{-1}$ or $C_2^{-1}$, which can be obtained through the performance measure ${\cal P}_1$ only, in either 
asymptotic limit of very fast
(shortest time scale $\tau_0$) or very slowly (largest $\tau_0$) decorrelating noise. From here on, we therefore 
focus on the angle performance measure ${\cal P}_1$ of Eq.~(\ref{eq:p1perf}) and consider the two asymptotic limits
in Eq.~(\ref{eq:p1}).

\begin{figure}[t]
 \begin{center}
\includegraphics[width=290px]{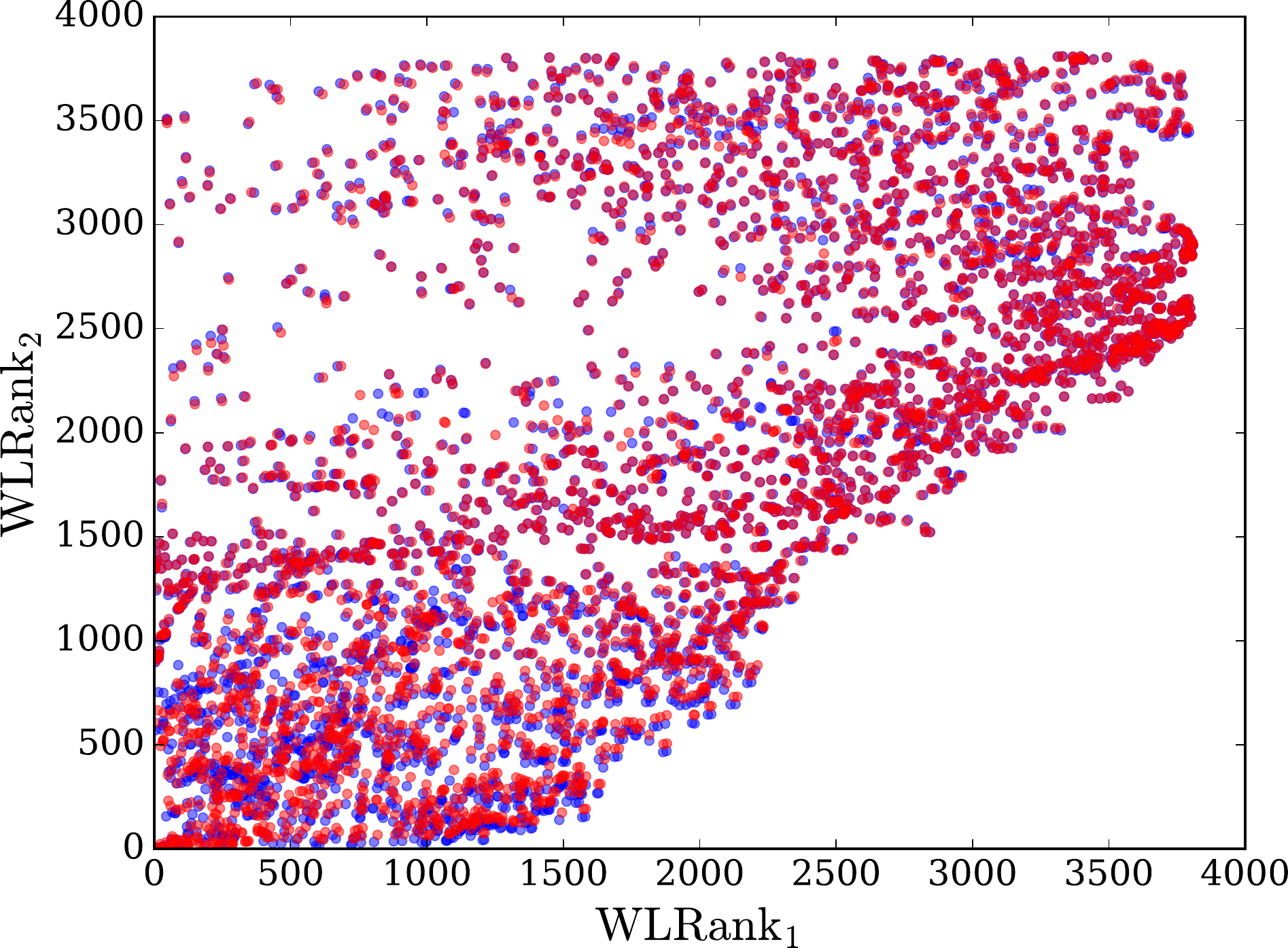}
  \caption{\footnotesize Comparison of the two nodal rankings WLRank$_1$ and WLRank$_2$ 
  obtained from 
  the generalized resistance centralities $C_1$ and $C_2$ respectively for the 3809 nodes of the
  European electric power grid sketched in Fig.~\ref{fig5}a (supplementary materials, materials and methods).
  Blue dots correspond to a
  moderate load during a standard winter weekday and red dots to a significantly heavier load corresponding to 
  the exceptional November 2016 situation with a rather large consumption and 
  twenty french nuclear reactors shut down.}
  \label{fig2}
 \end{center}
\end{figure}

We therefore define WLRank$_1$ and WLRank$_2$~\cite{caveat} as two rankings which order nodes 
from smallest to largest $C_1$ and $C_2$ respectively. 
Fig.~\ref{fig2} shows that they differ very significantly. In particular a number of nodes
are among the most critical according to WLRank$_1$ but not  to WLRank$_2$ and vice-versa. 
This discrepancy means that nodes are not central in an absolute sense, instead, their centrality and hence
how critical they are depends on details of the disturbance -- in the present case, the correlation time $\tau_0$ --
and the perfomance measure of interest. One should therefore chose to use one or the other 
centrality measure, according to the network sensitivity one wants to
check. 
%%%%%%%%%%%%%
%%%%%%%%%%%%%

%%%%%%
\begin{figure}[t]
 \begin{center}
\includegraphics[width=410px]{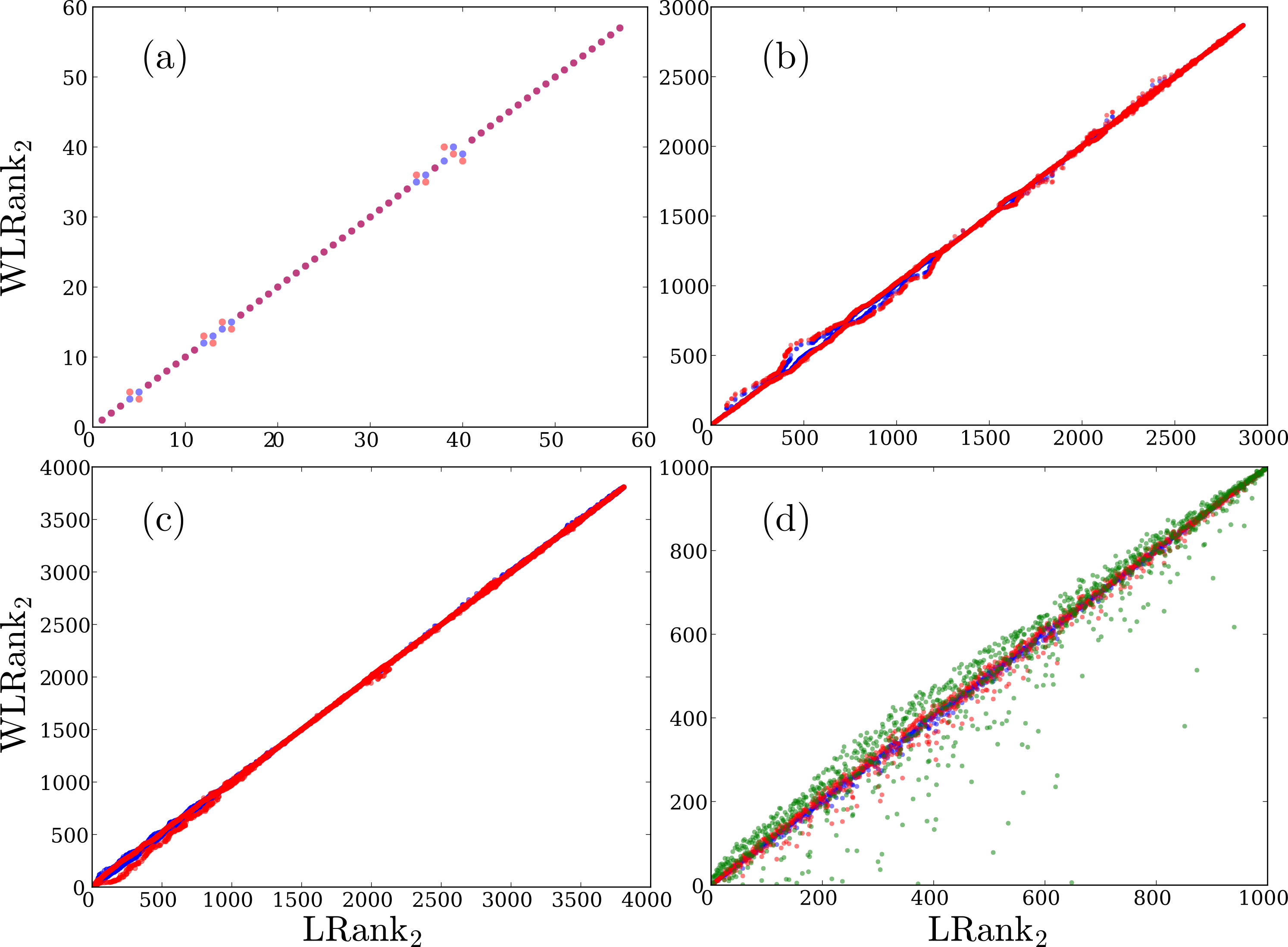}
  \caption{\footnotesize Comparison between LRank and WLRank corresponding to $\mathcal{P}_1$ for noisy disturbances with large 
  correlation time $\tau_0$. (a--c) Electric power grid models for normally (blue) and more heavily loaded (red) operating states
  governed by Eq.~(\ref{eq:swing}). 
  (a) IEEE 57 testcase where the more loaded case has   
  injections six times larger than the moderately loaded, tabulated case~\cite{IEEE}. (b) Pegase 2869 testcase where the more loaded case has injections 30\% larger than the moderately loaded, tabulated  case~\cite{MATP}. 
  (c) European electric power grid model sketched in Fig.~\ref{fig5}a 
  (supplementary materials, materials and methods) where the moderately loaded
  case corresponds to a standard winter weekday and the more heavily loaded case to the November 2016 situation with
  twenty french nuclear reactors offline. (d) Inertialess coupled oscillators governed by Eq.~(\ref{eq:swing}) 
  with $m_i=0$, $\forall i$, on a random network with 1000 nodes 
  obtained by  rewiring  a cyclic graph with constant nearest and next-to-nearest neighbor coupling with probability $0.5$
   (supplementary materials, materials and methods)~\cite{Wat98}. Natural frequencies are randomly distributed as
   $P_i \in [-1.8,1.63]$ (blue),  $P_i \in [-2.16,1.95]$ (red) and $P_i \in [-2.7,2.45]$ (green), corresponding to maximal angle 
   differences ${\rm max}(\Delta \theta) = 31^o$, $70^o$ and $106^o$ respectively. 
   } 
  \label{fig4}
 \end{center}
\end{figure}

The resistance centralities in Eqs.~(\ref{eq:p1p2}) correspond to the network defined by the weighted
Laplacian $\mathbb{L}(\bm \theta^{(0)})$ defined by Eq.~(\ref{eq:swing_lin}). They therefore depend on 
the unperturbed, operating state $\bm \theta^{(0)}$, consequently, WLRank depends not only on the 
nework topology, but also, as expected, on the natural frequencies and the coupling between the nodal degrees of 
freedom. As mentioned above, in the strong coupling limit, angle differences between coupled nodes remain small and 
$\mathbb{L}(\bm \theta^{(0)}) \simeq \mathbb{L}^{(0)}$. In that limit, one therefore expects nodal ranking to be 
given by resistance distances corresponding to the network Laplacian 
$\mathbb{L}^{(0)}$. How long this remains true is of central interest and
to answer this question 
we define further rankings LRank$_{1,2}$  as the rankings using resistance centralities $C_{1,2}^{(0)}$ obtained 
from the network Laplacian $\mathbb{L}^{(0)}$.
As long 
as angle differences between network-coupled nodes are not too large, the ranking LRank based on the 
network Laplacian matrix 
 is almost the same as the ranking true WLRank based on the weighted Laplacian. 
 This is shown in Fig.~\ref{fig4} for three electric power grid models and one random network of coupled oscillators. 
 For the electric power grid models, injections/natural frequencies are limited by the standard operational constraint 
 that the thermal limit of each power line
 is at most only weakly exceeded. This corresponds approximately to a maximal angle difference of 
 ${\rm max}(\Delta \theta) \simeq 30^o$ between any pair of coupled nodes. Accordingly, we find that 
 even in relatively strongly loaded
 power grids (corresponding for instance to the exceptional situation of the fall of 2016 when twenty french nuclear reactors were
 simultaneously offline; see red points in Fig.~\ref{fig4}c, there is not much of a difference between LRank and WLRank.  
 The two rankings start to differ from one another  
 only when at least some 
 natural frequencies become comparable with the corresponding nodal index, $P_i \lesssim \sum_j b_{ij}$, and angle 
 differences become very large. This case has been investigated 
 for an inertialess coupled oscillator system on a random rewired network with constant couplings 
 (supplementary materials, materials and methods)~\cite{Wat98}. It is shown in green in Fig.~\ref{fig4}d and 
 corresponds to ${\rm max}(\Delta \theta) = 106^o$.

\begin{figure}[t]
 \begin{center}
\includegraphics[width=290px]{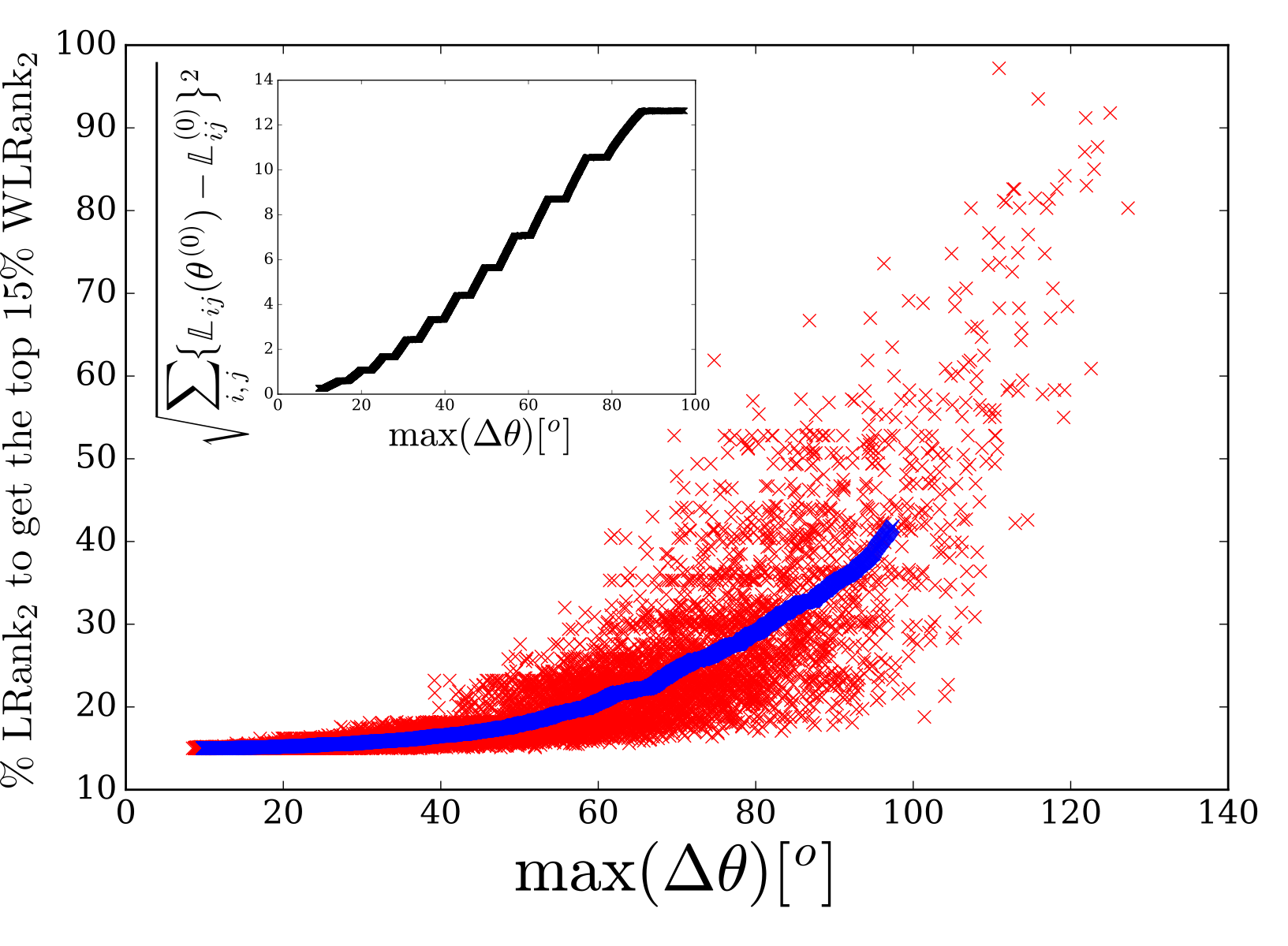}
\caption{\footnotesize Percentage of the nodes with highest LRank$_2$ necessary to give the top 15 \% ranked nodes with WLRank$_2$
for a random network of inertialess coupled oscillators with 1000 nodes 
  obtained by  rewiring  with probability $0.5$ a cyclic network with constant nearest and next-to-nearest neighbor coupling 
   (supplementary materials, materials and methods)~\cite{Wat98}. Each of the 12000 red crosses 
corresponds to one of 1000 random natural frequency vector $\bm P^{(0)}$ with components 
randomly distributed in $[-0.5,0.5]$ and 
summing to zero, multiplied by a prefactor $\beta = 0.4, 0.6, \ldots 2.4, 2.6$. The blue crosses correspond to
running averages over 500 red crosses with consecutive values of ${\rm max}(\Delta \theta)$. 
Inset : running averages of the Frobenius distance between 
the matrices $\mathbb{L}(\bm \theta^{(0)})$ and $\mathbb{L}^{(0)}$. The steps in the curve
reflect discrete increments  of $\beta$.}
  \label{fig3}
 \end{center}
\end{figure}

In Fig.~\ref{fig3} we investigate more closely when the approximate ranking LRank starts to differ from the true ranking WLRank.
To that end we used the randomly rewired model of inertialess coupled oscillators of Fig.~\ref{fig4}d and calculated the 
percentage of nodes with highest LRank$_2$ necessary to give the top 15 \% ranked nodes with WLRank$_2$.
The results are plotted as a function of the maximal angle difference between directly coupled nodes. 
Each of the 12000 red crosses 
in Fig.~\ref{fig3} corresponds to one of 1000 natural frequency vectors $\bm P^{(0)}$, with components 
randomly distributed in $[-0.5,0.5]$ and summing to zero, 
multiplied by a prefactor $\beta = 0.4, 0.6, \ldots 2.4, 2.6$. The blue crosses correspond to
running averages over 500 red crosses with consecutive values of ${\rm max}(\Delta \theta)$. 
One sees that, up to almost  ${\rm max}(\Delta \theta) \simeq 40^o$,
the set of the 18 \% of nodes with highest LRank$_2$ always includes the top 15 \% ranked nodes with WLRank$_2$. Similar 
results for obtaining the top 10 and 20 \% ranked nodes with WLRank$_2$, and for rankings using $C_1$ instead of $C_2$ 
are shown in the supplementary materials, materials and methods. 
 
That nodal ranking remains almost the same up to 
angle differences of about 40$^o$ is quite surprising, since coupling nonlinearities are already well developed there. 
This is illustrated in the inset of Fig.~\ref{fig3} which plots the  Frobenius distance
$ \sqrt{\sum_{ij} \left(\mathbb{L}_{ij}(\bm \theta^{(0)})-\mathbb{L}^{(0)}_{ij}\right)^2}$
between 
the network Laplacian $\mathbb{L}^{(0)}$ and the weighted Laplacian $\mathbb{L}(\bm \theta^{(0)})$. When 
${\rm max}(\Delta \theta) \simeq 40^o$, the Frobenius distance has already reached about 27 \% of its maximal
observed value, indicating that coupling nonlinearities are already significant. 
Yet, obtaining a desired set of the $n_s$ most critical nodes for any configuration with 
${\rm max}(\Delta \theta) \lesssim 40^o$, including cases with nonegligible nonlinearities, is achieved with 
a single matrix inversion of the network Laplacian $\mathbb{L}^{(0)}$, while
considering a slightly extended set of $n_s+ \delta n_s$ nodes with highest LRank, $\delta n_s/n_s \ll 1$.
This is a moderate price to pay, compared to the price of calculating WLRank
for each configuration, which each time requires inverting the weighted Laplacian matrix $\mathbb{L}(\bm \theta^{(0)})$.
That latter procedure would be too-time consuming for real-time assessment of large networks. 

\section{Conclusion}

We have formulated a key player problem in deterministic, network-coupled dynamical systems.  
The formulation is
based on the dynamical response to a nodal additive disturbance of the initial problem, and the most critical nodes
-- the key players -- are defined as those where the response to the disturbance is largest. 
While this manuscript focused on (i) noisy Ornstein-Uhlenbeck disturbances, 
(ii) network-coupled systems on undirected graphs, in particular with symmetric couplings $b_{ij}=b_{ji}$ in Eq.~(\ref{eq:swing}),
and (iii) 
performance measures of the  transient response  that are quadratic forms in the system's degrees of freedom,
the method is not restricted to such cases. First, it can be used to deal with different disturbances and in the 
supplementary materials, materials and methods, we calculate ${\cal P}_{1,2}$ for a box disturbance 
$\delta P_i(t) = \delta_{ik} \delta P_0 \Theta(t) \Theta(\tau_0-t)$ with the Heaviside function $\Theta(t)$. 
Remarkably, this disturbance gives the same ranking as the Ornstein-Uhlenbeck noise disturbance considered above.
Second, asymmetric couplings occurring e.g. in directed graphs~\cite{Del18}, in Kuramoto models with frustration~\cite{Ace05}
or in electric power grids with Ohmic dissipation~\cite{Bia08} can also be considered. In this case, 
the internodal coupling is given by asymmetric real matrices instead of symmetric Laplacian matrices.
However, the definition of the resistance distance, Eq.~(\ref{eq:rdistance}), 
remains valid even if $\mathbb{L}$ is replaced by an asymmetric matrix $\mathbb{A}$, 
in that it still gives $\Omega_{ii}=0$, 
$\Omega_{ij} \ge 0$, and $\Omega_{ij} \le \Omega_{ik}+\Omega_{ki}$, $\forall i,j,k$ as long as
the synchronous fixed point considered remains stable.
Third, nonquadratic performance measures can in principle be considered within the spectral decomposition used in this article.
One may think of average frequency nadir and rate of change of frequency, which 
are linear performance measures~\cite{Pag17,Guo18}. It is at present unclear whether these quantities can be 
analytically related to the
location of disturbances via resistance or other centralities. 

We gave an elegant answer to this key player problem :
ranking nodes from most to least critical is tantamount to ranking nodes from least to most central
in the sense of resistance centralities. Depending on how the problem is formulated -- mostly on 
details of the disturbance as well as on how the magnitude of the transient response is measured -- different centralities
have to be considered, giving different rankings. 
The key player problem in deterministic systems is therefore not uniquely defined
and its formulation must be tailored to reflect the most relevant dynamical properties one wants to
evaluate. Averaged rankings, reflecting several such properties simultaneously could also be considered. 

\section{Acknowledgments}
This work has been supported by the Swiss National Science Foundation under an AP Energy Grant.
We thank Robin Delabays and Tommaso Coletta for interesting discussions.

\bibliographystyle{apsrev}

\pagebreak
\begin{center}
\textbf{\large Supplementary Material for \\ The Key Player Problem in Complex Oscillator Networks and Electric Power Grids:  Resistance Centralities Identify Local Vulnerabilities}
\end{center}
%%%%%%%%%% Merge with supplemental materials %%%%%%%%%%
%%%%%%%%%% Prefix a "S" to all equations, figures, tables and reset the counter %%%%%%%%%%
\setcounter{equation}{0}
\setcounter{figure}{0}
\setcounter{table}{0}
\makeatletter
\renewcommand{\theequation}{S\arabic{equation}}
\renewcommand{\thefigure}{S\arabic{figure}}
\renewcommand{\bibnumfmt}[1]{[S#1]}
\renewcommand{\citenumfont}[1]{S#1}

\date{\today}

\maketitle
 
\section{Calculation of the Performance Measures}
We give some details of the calculation of the performance measures, Eqs.~(3) in the main text. These calculations 
generalize to second-order swing equations the results obtained for the first-order Kuramoto model in Ref.~\cite{SMTyl182}. Starting from Eq.~(1) in the main text, we consider a stable fixed-point solution ${\bm \theta}^{(0)}=(\theta_1^{(0)},\ldots ,\theta_n^{(0)}) $ with
unperturbed natural frequencies $\bm{P}^{(0)}$. We subject this state to a time-dependent disturbance
$\bm{P}(t) = \bm{P}^{(0)} + \delta \bm{P}(t)$, which makes angles become time-dependent, 
$\bm{\theta}(t) = \bm{\theta}^{(0)} + \delta \bm{\theta}(t)$. Linearizing the dynamics defined by Eq.~(1) of the main text about 
$\bm{\theta}^{(0)} $ and under the assumption that $d_i/m_i=\gamma$, $\forall i$, one obtains  
\begin{align}\label{eq:kuramoto_lin}
\delta \ddot{\bar{\bm \theta}} + \gamma\delta \dot{\bar{\bm \theta}} &= {\bm M}^{-1/2}\delta {\bm P} - {\bm M}^{-1/2}{\mathbb L}(\bm\theta^{(0)}){\bm M}^{-1/2} \, \delta {\bar{\bm \theta}} \, ,
\end{align}
where we introduced matrices with elements 
$D_{ij}=\delta_{ij} \, d_i=\gamma M_{ij}$ and new angle coordinates $\delta {\bar{\bm \theta}}={\bm M}^{1/2}\delta {\bm \theta}$. 
The weighted Laplacian matrix ${\mathbb L}(\bm \theta^{(0)} )$ is defined as
\begin{equation}\label{eq:laplacian}
{\mathbb L}_{ij} = 
\left\{ 
\begin{array}{cc}
-b_{ij} \cos(\theta_i^{(0)} - \theta_j^{(0)}) \, , & i \ne j \, , \\
\sum_k b_{ik} \cos(\theta_i^{(0)} - \theta_k^{(0)}) \, , & i=j \, .
\end{array}
\right.
\end{equation}
This Laplacian is minus the stability matrix of the linearized dynamics about a stable synchronous state. It
is therefore positive semidefinite, with its largest eigenvalue $\lambda_1=0$ corresponding to a constant
eigenvector ${\bf u}_1=(1,1,1,...1)/\sqrt{n}$, and $\lambda_\alpha>0$, $\alpha=2,3,...n$. We define the matrix ${\mathbb L}^{M}={\bm M}^{-1/2}{\mathbb L}{\bm M}^{-1/2}$ with eigenvectors ${\bf u}^{M}_{\alpha}$ and eigenvalues $\lambda^{M}_{\alpha}$, for $\alpha=1,2,...n$. 
To calculate the response of the system to $\delta {\bm P}(t)$, we expand angle deviations over the eigenstates ${\bf u}^M_\alpha$ of ${\mathbb L}^M$,
$ \delta \bar{\bm{\theta}}(t)=\sum_\alpha c_\alpha(t) \, {\bf u}^{M}_\alpha$. Eq.~(\ref{eq:kuramoto_lin}) becomes 
\begin{equation}\label{eq:kuramoto_ca}
\ddot{c}_\alpha(t)+\gamma\dot{c}_\alpha(t) = {\bm M}^{-1/2}\delta \bm{P}(t) \cdot {\bf u}^M_\alpha - \lambda^M_\alpha c_\alpha(t) \, .
\end{equation}
The disturbance starts at $t=0$ and therefore $\delta \bar{\bm \theta}(0)=0$ and $\delta \dot{\bar{\bm \theta}}(0)=0$.
Performing a Laplace transform on Eq.~\eqref{eq:kuramoto_ca}, one gets
\begin{eqnarray}
s^2c_\alpha(s)+\gamma \, s \, c_\alpha(s)=\lambda_\alpha^M c_\alpha(s)+ ({\bm M}^{-1/2}\delta{\bm P}\cdot {\bf u}_{\alpha}^M)(s) \; ,
\end{eqnarray}
where $c_\alpha(s)=\int_0^{t}e^{-st'}c_\alpha(t')dt'$ and 
$ ({\bm M}^{-1/2}\delta{\bm P}\cdot {\bf u}_{\alpha}^M)(s)=\int_0^te^{-st'}{\bm M}^{-1/2}\delta{\bm P}(t')\cdot {\bf u}_{\alpha}^M \, dt'$. Finally one obtains the Laplace transformed expansion coefficients 
of the angles over the eigenbasis of ${\bf u}^M_\alpha$ of ${\mathbb L}^M$,
\begin{eqnarray}
c_\alpha(s)=({\bm M}^{-1/2}\delta{\bm P}\cdot {\bf u}_{\alpha}^M)(s) \Big / \left(s-\frac{-\gamma+\Gamma_{\alpha}}{2}\right)\left(s+\frac{\gamma + \Gamma_\alpha}{2}\right) \; ,
\end{eqnarray}
with $\Gamma_\alpha=\sqrt{\gamma^2-4\lambda_\alpha^M}$. Applying an inverse Laplace transform leads to,
\begin{eqnarray}\label{eq:calpha}
c_{\alpha}(t)=e^{\frac{-\gamma-\Gamma_{\alpha}}{2}t}\int_0^{t}e^{{\Gamma_{\alpha}}t'}\int_{0}^{t'}{\bm M}^{-1/2}\delta {\bm{P}}(t'')\cdot {\bf{u}}^M_{\alpha}e^{\frac{\gamma-\Gamma_{\alpha}}{2}t''}dt''dt' \;.
\end{eqnarray}
The time-dependence of angle and frequency degrees of freedom  is then given by,
\begin{eqnarray}
\delta {{\bm \theta}}(t)&=&{\bm M}^{-1/2}\delta \bar{{\bm \theta}}(t)=\sum_{\alpha}c_\alpha(t) {\bm M}^{-1/2}{\bf u}_\alpha^M \; , \\
\delta \dot{{\bm \theta}}(t)&=&{\bm M}^{-1/2}\delta \dot{\bar{{\bm \theta}}}(t)=\sum_{\alpha}\dot{c}_\alpha(t) {\bm M}^{-1/2}{\bf u}_\alpha^M \; .
\end{eqnarray}
%The variances of the angle displacements $\delta {{\bm \theta}}(t)$ and frequencies deviations $\delta \dot{{\bm \theta}}(t)$ are then given by,
The variances $p_1(t)$ and $p_2(t)$ of the angle and frequency deviations read,
\begin{eqnarray}
p_1(t)=\delta {{\bm \theta}}^2(t)=\sum_{\alpha,\beta}c_\alpha(t)c_\beta(t) {{\bf u}_{\beta}^{M}}^{\top}{\bm M}^{-1}{\bf u}_{\alpha}^{M} \; , \\
p_2(t)=\delta \dot{{\bm \theta}}^2(t)=\sum_{\alpha,\beta}\dot{c}_\alpha(t)\dot{c}_\beta(t) {{\bf u}_{\beta}^{M}}^{\top}{\bm M}^{-1}{\bf u}_{\alpha}^{M} \; .
\end{eqnarray}
When $d_i=d=\gamma m_i$ $\forall i$, both matrices ${\mathbb L}$ and ${\mathbb L}^M$ have the same eigenvectors and $\lambda_\alpha^M=\lambda_{\alpha}/m$. We assume homogeneous inertia and damping factor for the calculations in the 
next two paragraphs.

{\bf Correlated Noisy disturbances}

In the case of stochastic disturbances that persist in time, we average the $p_i$'s as follows,
\begin{eqnarray}\label{eq:PT}
\mathcal{P}_i=\lim_{T\rightarrow \infty} T^{-1}\int_0^T \overline{p_i(t)} dt  \; \;,\, i=1,2 \, ,
\end{eqnarray}
where $\overline{p_i(t)}$  indicates an average taken over the ensemble defined by e.g. the moments of the stochastic disturbance.
We consider Ornstein-Uhlenbeck 
 correlated noise on a single node, $k$, with zero mean $\overline{\delta P_k(t)}=0$ and second moment $\overline{\delta P_i(t_1)\delta P_j(t_2)}=\delta_{ik} \delta_{jk} \, \delta P_{0}^2\exp[-|t_1-t_2|/\tau_0 ]$, correlated over a typical time scale $\tau_0$ and uniform inertia and damping. We have,
\begin{eqnarray}
\mathcal{P}_1&=&\lim_{T\rightarrow \infty}T^{-1} \sum_{\alpha \ge 2}\int_0^T \overline{c_\alpha^2(t)} dt\\
&=&\lim_{T\rightarrow \infty} T^{-1}\sum_{\alpha \ge 2}\int_0^T e^{-(\gamma + \Gamma_\alpha)t}\int_0^t\int_0^t e^{\Gamma_\alpha(t_1'+t_2')}\times\\
&&\int_0^{t_1'}\int_0^{t_2'}\sum_{i,j}\frac{u_{\alpha,i} u_{\alpha,j}}{m}\, \overline{\delta P_i(t_1'') \delta P_j(t_2'')} \, e^{\frac{\gamma-\Gamma_\alpha}{2}(t_1''+t_2'')} \, dtdt_1'dt_2'dt_1''dt_2'' \; . \nonumber
\end{eqnarray}
For homogeneous damping and inertia one has 
$\Gamma_\alpha=\sqrt{\gamma^2-4\lambda_\alpha/m}$. The integrals can be performed  straightforwardly and
one obtains
%\begin{eqnarray}
%\mathcal{P}_3&=&\frac{\delta P_0^2}{m_k} \sum_{\alpha\ge 2}\frac{{u_{\alpha,k}^M}^2}{(\lambda_{\alpha}^M\tau_0 + \gamma + \tau_0^{-1})} \; .
%\end{eqnarray}
%\begin{eqnarray}
%\mathcal{P}_3&=&\lim_{T\rightarrow \infty}\frac{\gamma}{T} \sum_{\alpha \ge 2}\int_0^T \overline{c_\alpha^2(t)} dt\\
%&=&\lim_{T\rightarrow \infty} \frac{\gamma}{T}\sum_{\alpha \ge 2}\int_0^T e^{-(\gamma + \Gamma_\alpha)t}\int_0^t\int_0^t e^{\Gamma_\alpha(t_1'+t_2')}\times\\
%&&\int_0^{t_1'}\int_0^{t_2'}\sum_{i,j}\frac{u_{\alpha,i}^M u_{\alpha,j}^M}{(m_i m_j)^{1/2}}\overline{\delta P_i(t_1'') \delta P_j(t_2'')} e^{\frac{\gamma-\Gamma_\alpha}{2}(t_1''+t_2'')}dtdt_1'dt_2'dt_1''dt_2'' \; . \nonumber
%\end{eqnarray}
%After some algebra we get,
%\begin{eqnarray}
%\mathcal{P}_3&=&\frac{\delta P_0^2}{m_k} \sum_{\alpha\ge 2}\frac{{u_{\alpha,k}^M}^2}{(\lambda_{\alpha}^M\tau_0 + \gamma + \tau_0^{-1})} \; .
%\end{eqnarray}
%With the additional constraint of uniform damping and inertias, i.e. $d_i=d=\gamma m_i$ $\forall i$, $\mathbb{L}$ and $\mathbb{L}^M$ have the same eigenvectors i.e. ${\bf u}_\alpha={\bf u}_\alpha^M$ and eigenvalues such that $\lambda_\alpha^M=\lambda_\alpha/m$ and we have,
\begin{subequations}\label{eqs:P1P2}
\begin{align}
\mathcal{P}_1&=\delta P_0^2\sum_{\alpha \ge 2}\frac{u_{\alpha,k}^2(\tau_0 + m/d)}{\lambda_\alpha(\lambda_\alpha\tau_0 + d + m\tau_0^{-1})} \; , \\ 
\mathcal{P}_2&=\delta P_0^2\sum_{\alpha \ge 2}\frac{u_{\alpha,k}^2}{d(\lambda_\alpha\tau_0 + d + m\tau_0^{-1})} \; .
\end{align}
\end{subequations}
Taking the two limits $\lambda_\alpha \tau_0 \gg d$, $\lambda_\alpha \tau_0^2 \gg m$ and $\lambda_\alpha \tau_0 \ll d$, 
$\lambda_\alpha \tau_0^2 \ll m$,
Eqs.~(6a,b) of the main text are then easily obtained. Note that the above computation can be done relaxing the uniform inertia and damping hypothesis. The performance measures for Kuramoto oscillators are obtained for $m=0$ \cite{SMTyl182}. The asymptotics are then obtained by taking the asymptotic limits of large/small $\tau_0$ only after setting $m=0$. One obtains,
\begin{subequations}\label{eqSM:p1p2}
\begin{equation}\label{eqSM:p1}
\mathcal{P}_1 = \left\{
  \begin{array}{lr}
    \big(\delta P_0^2\tau_0 \big)\big/ d ) \left({C_1^{-1}(k)}-n^{-2}\Kf_1\right) \; \; ,\;  \lambda_\alpha\tau_0\ll 1 \, , \\
    \delta P_0^2\left( {C_2^{-1}(k)}-n^{-2}\Kf_2\right) \; \; ,\;  
    \lambda_\alpha \tau_0 \gg d  \, , 
  \end{array}
\right.
\end{equation}
\begin{equation}
\mathcal{P}_2 = \left\{
  \begin{array}{lr}
    \big(\delta P_0^2\tau_0 \big/ d \big) \big( n-1 \big) \big/ n   \; \;  ,\; \lambda_\alpha\tau_0\ll 1 \, , \\
    \big(\delta P_0^2 \big/ d\tau_0 \big) \left( {C_1^{-1}(k)}-n^{-2}\Kf_1\right) \; \; ,\;  \lambda_\alpha \tau_0 \gg d\, ,\label{eqSM:p2}
  \end{array}
\right.
\end{equation}
\end{subequations}
where we use the generalized resistance centralities $C_{1,2}(i)$ and Kirchhoff indices $\Kf_{1,2}$ 
discussed in Section~\ref{rdki} below.

{\bf Box disturbances}

The same kind of computation as for the noisy disturbance can be done with a box disturbance acting on node $k$, i.e. $\delta P_i(t)=\delta_{ik} \, \delta {P}_{0} \, \Theta(t)\Theta(\tau_0-t)$ with the Heaviside step function $\Theta(t)=0$ for $t<0$ and $\Theta(t)=1$ for $t\ge 1$. As the perturbation is limited in time, we consider the performance measures,
\begin{eqnarray}
\mathcal{P}^\infty_1=\sum_i\int_0^\infty |\delta \theta_i -\Delta(t)|^2 dt   \; , \\
\mathcal{P}^\infty_2=\sum_i\int_0^\infty |\delta \dot{\theta}_i -\dot{\Delta}(t)|^2 dt   \; ,
\end{eqnarray}
instead of (\ref{eq:PT}).
For uniform inertia and damping one obtains, 
\begin{eqnarray*}
\mathcal{P}^\infty_1 &=&\frac{\delta P_0^2m}{8\gamma} \sum_{\alpha\ge 2}\frac{{u_{\alpha ,k}^2}}{\Gamma_{\alpha}\lambda_{\alpha}^3}\left[ 2\Gamma_{\alpha} ( 4\gamma\tau_0\lambda_\alpha /m - 3\gamma^2 - \Gamma_\alpha^2 ) + ( \gamma+\Gamma_\alpha )^3e^{-\tau_0 \frac{(\gamma-\Gamma_\alpha)}{2}} - ( \gamma-\Gamma_\alpha )^3e^{-\tau_0\frac{(\gamma+\Gamma_\alpha)}{2}} \right] \, , \\
\mathcal{P}^\infty_2&=& \frac{\delta P_0^2}{2d}\sum_{\alpha\ge 2}\frac{{u_{\alpha ,k}^2}}{\Gamma_\alpha\lambda_\alpha}\left[2\Gamma_\alpha -(\gamma+\Gamma_\alpha)e^{-\frac{\tau_0(\gamma-\Gamma_\alpha)}{2}} +(\gamma-\Gamma_\alpha)e^{-\frac{\tau_0(\gamma+\Gamma_\alpha)}{2}}  \right] \; ,
\end{eqnarray*}
with $\Gamma_\alpha=\sqrt{\gamma^2-4\lambda_\alpha/m}$.
The two asymptotic limits of  large and small $\tau_0$ are given by,
\begin{subequations}
\begin{equation}
\mathcal{P}^\infty_1 = \left\{
  \begin{array}{lr}
    \big(\delta P_0^2 \tau_0^2 \big/ 2d ) \left({C_1^{-1}(k)}-n^{-2}\Kf_1\right) \; \; ,\;  (\gamma \pm \Gamma_\alpha) \tau_0\ll 1 \, ,  \\
    \delta P_0^2\tau_0\left( {C_2^{-1}(k)}-n^{-2}\Kf_2\right) \; \; ,\;   (\gamma \pm \Gamma_\alpha) \tau_0\gg 1 \; {\rm and} \; \lambda_\alpha \tau_0/d \gg 1 \, , \; 
  \end{array}
\right.
\end{equation}
\begin{equation}
\mathcal{P}^\infty_2 = \left\{
  \begin{array}{lr}
    \big(\delta P_0^2\tau_0^2 \big/ 2md \big) \big( n-1 \big)\big/ n   \; \;  ,\;   (\gamma \pm \Gamma_\alpha) \tau_0\ll 1 \, , \\
    \big(\delta P_0^2 \big/ d \big) \left( {C_1^{-1}(k)}-n^{-2}\Kf_1\right) \; \; ,\;   (\gamma \pm \Gamma_\alpha) \tau_0\gg 1 \, ,
  \end{array}
\right.
\end{equation}
\end{subequations}
which are also given by resistance centralities and Kirchhoff indices. 
\section{Resistance Distances, Centralities and Kirchhoff Indices}\label{rdki}
The resistance centralities $C_1$ and $C_2$ can be expressed as functions of the distribution of resistance distances $\Omega_{ij}$, between any pairs of nodes $(i,j)$ of the network. The Laplacian matrix $\mathbb{L}$ of the network has one zero eigenvalue associated to the constant eigenvector ${u}_{1,i}=1/\sqrt{n}$, its pseudoinverse $\mathbb{L}^{\dagger}$ is defined by \cite{SMKle932},
\begin{eqnarray}
{\mathbb{L}}{{\mathbb{L}}^{\dagger}}={\mathbb{L}^{\dagger}}{{\mathbb{L}}}=\mathbb{I}-{\bf u}_1^{\top}{\bf u}_1 \; ,
\end{eqnarray}
from which the resistance distance between nodes $i$ and $j$ is expressed as,
\begin{eqnarray}\label{eq:rd}
\Omega_{ij}={\mathbb{L}}^{\dagger}_{ii} + {\mathbb{L}}^{\dagger}_{jj} - {\mathbb{L}}^{\dagger}_{ij} -{\mathbb{L}}^{\dagger}_{ji} \; .
\end{eqnarray}
Using the eigenvectors of $\mathbb{L}$ we can rewrite Eq.~(\ref{eq:rd}) as~\cite{SMTyl182},
\begin{eqnarray}\label{eq:RD}
\Omega_{ij}=\sum_{\alpha \ge 2}\frac{(u_{\alpha,i}-u_{\alpha,j})^2}{\lambda_\alpha} \; .
\end{eqnarray} 

The resistance distance is a graph metric in the sense that : i) $\Omega_{ii}=0$, $\forall i$, ii) $\Omega_{ij}\ge 0$, $\forall i,j$,  and iii) $\Omega_{ij}+\Omega_{jk}\ge \Omega_{ik}$, $\forall i,j,k$ (triangle inequality) \cite{SMKle932}. The Kirchhoff index of a network is obtained from the resistance distances by summing over all pairs of nodes, \cite{SMKle932}
\begin{eqnarray}\label{eq:KF}
\Kf_1= \sum_{i<j}\Omega_{ij}=n\sum_{\alpha \ge 2}\lambda_\alpha^{-1} \; .
\end{eqnarray}
The Kirchhoff index is, up to a normalization factor, the mean resistance distance over the whole graph.

We generalize this definition of the resistance distance for matrices that are powers of the original Laplacian matrix, 
$\mathbb{L}'=\mathbb{L}^p$ and thus
$\left[\mathbb{L}'\right]^{\dagger}=\left[{\mathbb{L}^p } + {\bf u}_1^{\top}{\bf u}_1\right]^{-1}$. One has
\begin{eqnarray}
\Omega_{ij}^{(p)}=[\mathbb{L}_{ii}']^{\dagger} + [\mathbb{L}_{jj}']^{\dagger} - [\mathbb{L}_{ij}']^{\dagger}- [\mathbb{L}_{ji}']^{\dagger} \; .
\end{eqnarray}
The eigenvectors of $\mathbb{L}'$ are the same as those of $\mathbb{L}$. Thus we have,
\begin{eqnarray}
\Omega_{ij}^{(p)}=\sum_{\alpha \ge 2}\frac{(u_{\alpha,i}-u_{\alpha,j})^2}{\lambda_\alpha^p} \; .
\end{eqnarray} 
We still have to check that the generalized resistance distances $\Omega_{ij}^{(p)}$ have the three properties of a graph metric. We remark that $\Omega_{ij}^{(p)}$ corresponds to the resistance distance between nodes $i$ and $j$ in a new graph whose Laplacian is $\mathbb{L}'=\mathbb{L}^p$. Therefore it is sufficient to show that $\mathbb{L}'$ is also a Laplacian matrix. to that end
we demonstrate that the product of two Laplacian matrices $\mathcal{A}$ and $\mathcal{B}$ is still a Laplacian matrix. For a Laplacian matrix
$\mathcal{A}$ one has (i) $\sum_i {\mathcal{A}}_{ij}=0$, (ii) ${\mathcal{A}}_{ii}=-\sum_{j\neq i}\mathcal{A}_{ij}$. From these generic
properties of Laplacian matrices, 
matrix elements of 
the product $\mathcal{A} \mathcal{B}$ satisfy
\begin{eqnarray}
\sum_{j}[\mathcal{A}\mathcal{B}]_{ij}&=&\sum_{j,k}\mathcal{A}_{ik}\mathcal{B}_{kj}=0 \; , \\
\sum_{j\neq i}[\mathcal{A}\mathcal{B}]_{ij}&=&\sum_{j}[\mathcal{A}\mathcal{B}]_{ij}- [\mathcal{A}\mathcal{B}]_{ii}=-[\mathcal{A}\mathcal{B}]_{ii} \; .
\end{eqnarray}
We conclude that the product $\mathcal{A} \mathcal{B}$ is also a Laplacian matrix, and therefore, 
 the generalized resistance distances $\Omega_{ij}^{(p)}$ have the three properties of a graph metric. With the generalized resistance distances, we can define generalized Kirchhoff indices \cite{SMTyl182},
 \begin{eqnarray}\label{eq:GKF}
\Kf_p= \sum_{i<j}\Omega_{ij}^{(p)}=n\sum_{\alpha \ge 2}\lambda_\alpha^{-p} \; .
\end{eqnarray}

%Note that the generalized resistance distance $\Omega_{ij}^{(p)}$ correspond to the resistance distance between nodes $i$ and $j$ in a graph defined by the Laplacian matrix $\mathbb{L}^p$.
The relation between the resistive centrality $C_1(i)$ and the resistance distance is obtained from Eqs.~(\ref{eq:RD}) and (\ref{eq:KF}),
\begin{eqnarray}
C_1(i)&=&\left[n^{-1}\sum_{j}\Omega_{ij}\right]^{-1}=\left[ \sum_{\alpha\ge 2}\frac{u_{\alpha,i}^2}{\lambda_\alpha} + n^{-2}\Kf_1 \right]^{-1} \; .
\end{eqnarray}
The expression for $C_2(i)$ involves higher moments of the distribution of resistance distances.
We obtain
\begin{eqnarray*}
C_2(i)
%&=&\left[n^{-1}\sum_{j}\Omega_{ij}^{(2)}\right]^{-1}=\left[\sum_{\alpha\ge 2}\frac{u_{\alpha,i}^2}{\lambda_\alpha^2} + n^{-2}\Kf_2 \right]^{-1} \\
%4\frac{u_{\alpha,i}^2}{\lambda_\alpha^2}&=& \sum_{j}\Omega_{ij}^2-n^{-1}\left[ \sum_{j}\Omega_{ij} \right]^2  \\
%&+&2 \sum_j \left( \Omega_{ij} - \left[ n^{-1}\sum_k (\Omega_{ik} + \Omega_{jk}) -2n^{-2}\Kf_1 \right]\right)\left(n^{-1}\sum_k\Omega_{jk}-n^{-2}\Kf_1 \right) \\
%&-&  \sum_j \left[n^{-1}\sum_k \Omega_{jk}-n^{-2}\Kf_1 \right]^2 + n^{-3}\Kf_1 \\
&=&\sum_{j}\Omega_{ij}^2 - n\; C_1^{-2}(i) + 2\sum_j \Omega_{ij} \;{C_1^{-1}(j)} -4 \;{C_1^{-1}(i)}\; n^{-1}\Kf_1 -3\sum_j {C_1^{-2}(j)} + 12n^{-3}\Kf_1^2  \; .
\end{eqnarray*}

\section{Numerical Models}

We checked our analytical results against numerical ones obtained for four different models which we briefly describe here. 

\subsection{European electric power grid}

We have constructed a model of the European high voltage electrical grid. It is 
composed of $3809$ consumer and generator nodes
connected to one another by $4944$ lines. The geographic location of each node and the location of the lines between them
has been extracted from the
ENTSO-E database \cite{SMWie16}. Line capacities $b_{ij}$ between nodes 
have been normalized proportionally to the inverse of their length. 
The operational states (injections and consumptions) of the power grid are obtained via an optimal power flow
which constrains the load flows on each line with the thermal limit of the latter and takes into account technical specificities for each power plant~\cite{SMMATP,Pag18}. 
The two operational states considered in Figs.~3 and 4c of the main text correspond to a typical electric power  
consumption situation 
in winter (blue) and a case reproducing the extraordinary situation of November 2016, with a relatively high power demand and twenty french nuclear reactors offline (red). For the numerical simulations in Fig.~1 of the main text, we used the first case. 
For this model, 
the network Laplacian matrix has a spectrum distributed in the interval $\lambda_\alpha \in [0.0458, 26678.4395]$
(in the per unit system~\cite{SMBia08}).

\subsection{IEEE 57 bus test case}

The IEEE 57 bus test case is a standardly used model of an electric power grid~\cite{SMIEEE}. It
is composed of $57$ buses including $7$ generators and  $80$ lines.
In Fig.4a of the main text, we use the tabulated operational state as well as a state where the tabulated 
loads are increased by a factor six~\cite{SMIEEE}.
The spectrum of the Laplacian is distributed in the interval  $\lambda_\alpha\in [0.2796,118.6186]$ (in the per unit system~\cite{SMBia08}).

\begin{figure}
\includegraphics[width=0.99\textwidth]{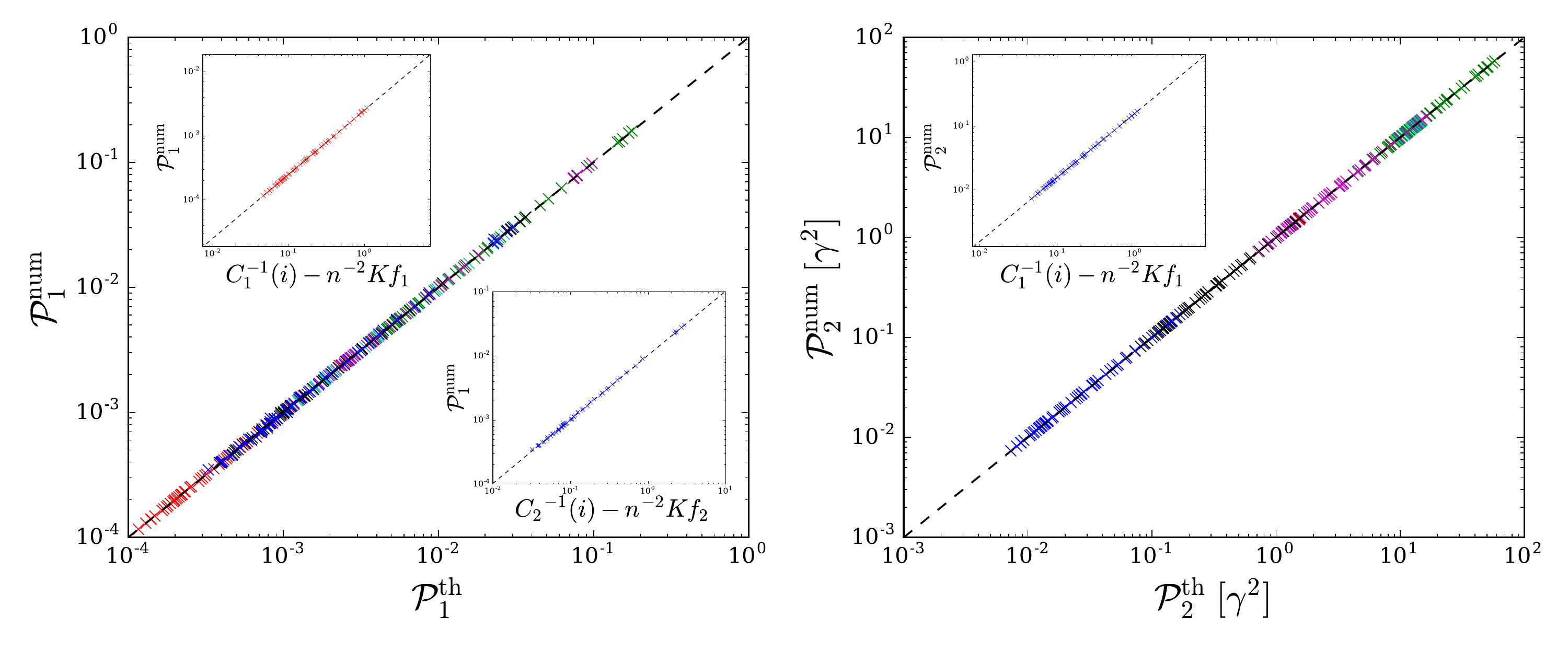}
\caption{Comparison between theoretical predictions and numerical results for both performance measures
  ${\cal P}_1$ and ${\cal P}_2$. 
  Each point corresponds to a noisy disturbance 
  on a single node of the IEEE 57 bus test case~\cite{SMIEEE} with magnitude $\delta P_0=0.1$ and correlation 
  times $\gamma \tau_0=4\cdot 10^{-4}$ (red crosses), $4\cdot 10^{-3}$ (cyan), $4\cdot 10^{-2}$ (green), $4\cdot 10^{-1}$ (purple),  
  $4$ (black) and $40$ (blue). Time scales are defined by the ratio of damping to inertia coefficients
  $\gamma=d_i/m_i=0.4 s^{-1}$ which is assumed constant with $d_i=0.004s$.
  The insets show ${\cal P}_1$ and ${\cal P}_2$ as a function of the resistance distance-based 
  graph-theoretic predictions of Eqs.~(5) in the main text, valid in both limits of very large and very short noise decorrelation time
  $\tau_0$. Not shown is the limit of short $\tau_0$ for ${\cal P}_2$, which gives a node-independent
  result.}\label{fig:S2} 
\end{figure}

Fig.~\ref{fig:S2} shows data similar to Fig.~1 in the main text for the IEEE 57 bus test case. The insets shows the asymptotic limits 
of very large and very small $\tau_0$, where ${\cal P}_{1,2}$ are predicted to be 
linear functions of the resistance centralities $C_{1,2}$ (see main text).

\subsection{MATPOWER Pegase 2869 Test Case}

\begin{figure}
\includegraphics[width=0.99\textwidth]{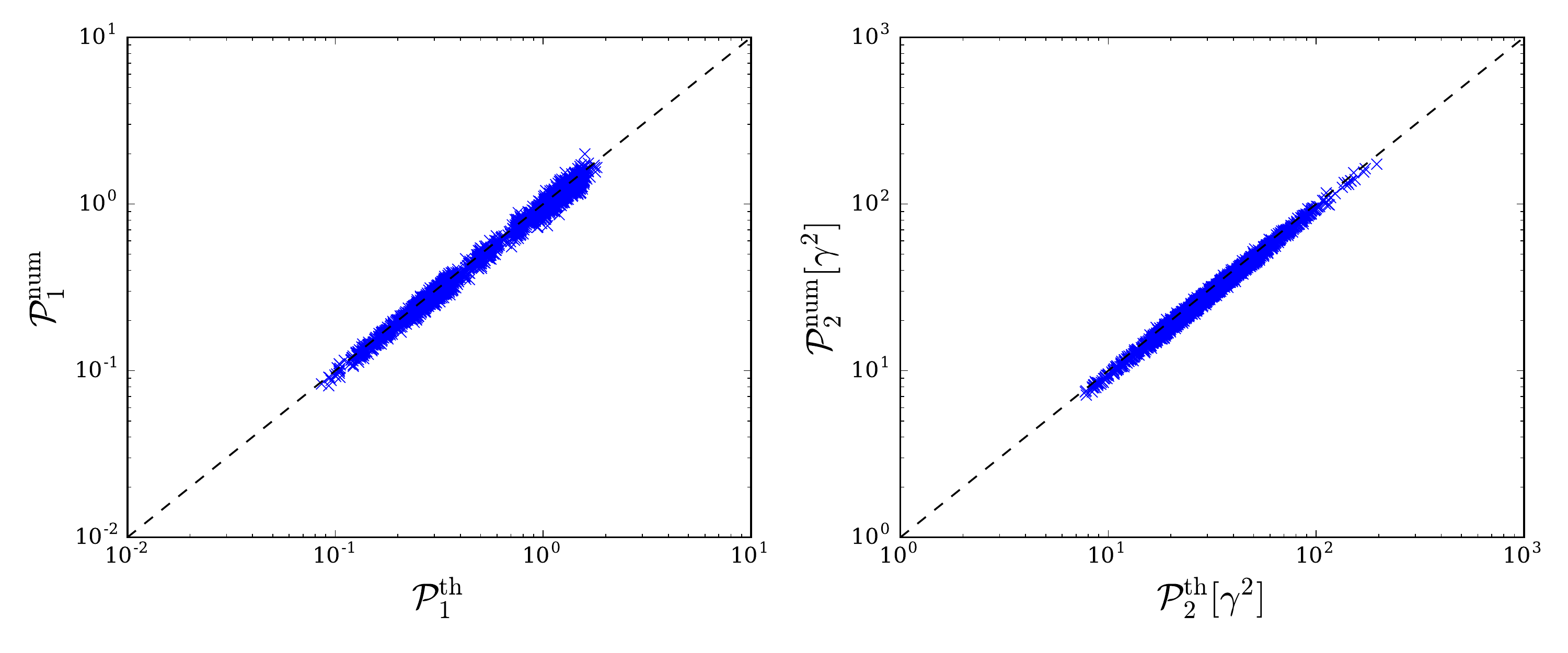}
\caption{Comparison of the performance measures $\mathcal{P}_1$, $\mathcal{P}_2$ obtained numerically and Eqs.~(\ref{eqs:P1P2}). Each point corresponds to a noisy disturbance on a single node of the Pegase 2869 test case \cite{SMMATP} with magnitude $\delta P_0=0.1$ and correlation time $\gamma \tau_0 = 0.4$ and ratio of damping to inertia $\gamma=0.4 s^{-1}$ with $d_i=0.007s$.}\label{fig:S6}
\end{figure}

The MATPOWER 
test case Pegase 2869 is a model representing a part of the European high voltage transmission grid~\cite{SMMATP}. It is composed of $2869$ buses including $510$ generators and $4582$ lines. In Fig.4b of the main text, 
we use the tabulated operational state as well as a state where injections are 30\% larger~\cite{SMMATP}.The spectrum of the Laplacian is distributed in the interval $\lambda_\alpha\in [0.03536,27156.901]$ (in the per unit system~\cite{SMBia08}). 
Fig.~\ref{fig:S6} shows data similar to Fig.~1 in the main text for this model.

\subsection{Random Network}

We finally used a random network obtained by random rewiring of edges with probability $0.5$ of a single-cycle network
with 1000 nodes
with nearest and next-to-nearest couplings~\cite{SMWat982}.
Edges have the same weight $b_{ij}=b_0=1s^{-1}$.
The spectrum of the Laplacian is distributed in the interval $\lambda_{\alpha}\in [0.39b_0,10.47b_0]$.

In our numerics, we define a first-order, inertialess Kuramoto model on this random network. Fig.4d of the main text considers
various distribution of natural frequencies, including one (green) which is close to instability with angle differences larger
than 90$^o$.

\section{Numerical Comparison of LRank with WLRank}

\begin{figure}
\includegraphics[width=0.99\textwidth]{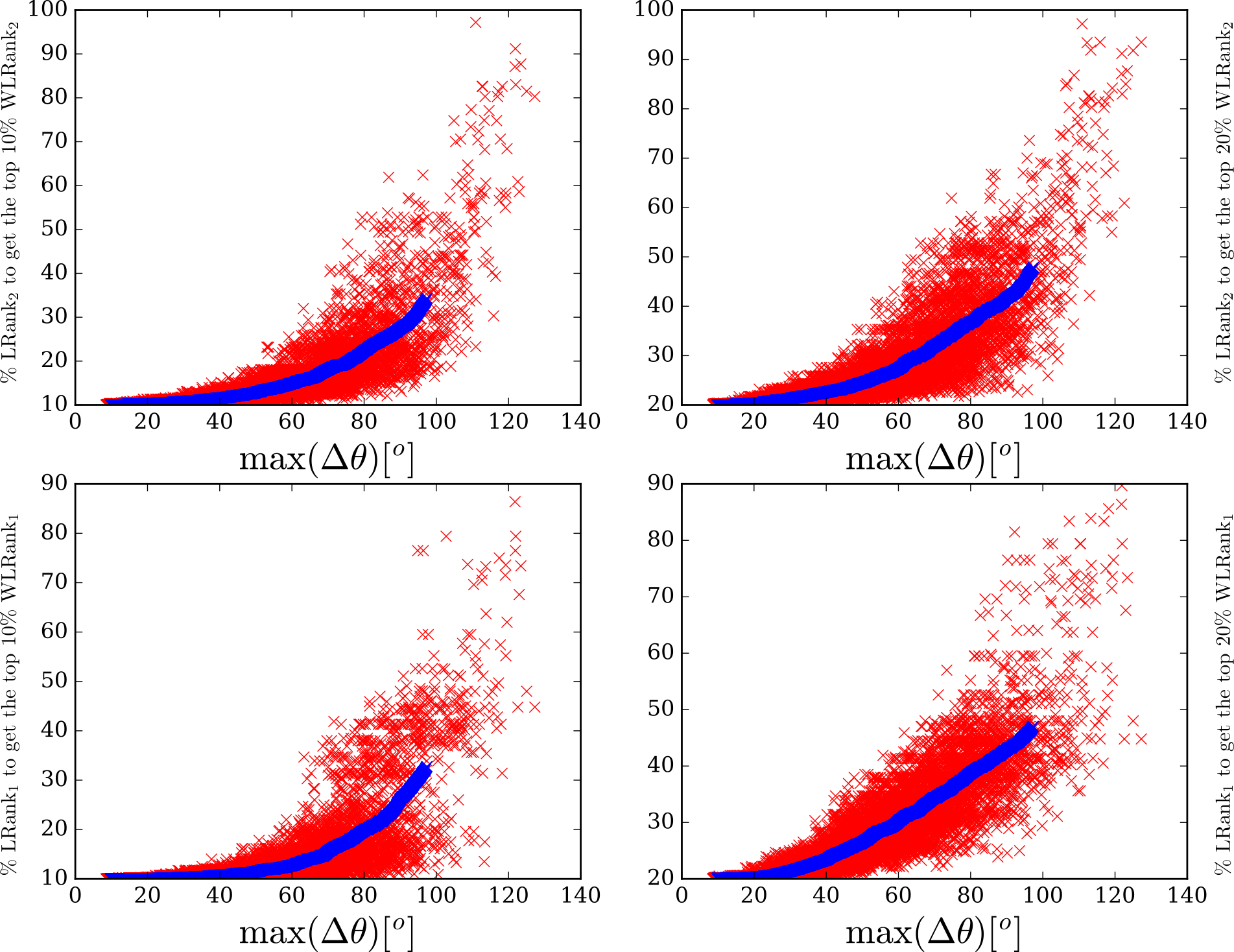}
\caption{\footnotesize Percentage of the nodes with highest LRank$_{1,2}$ necessary to give the top 10 \% (left), 20\% (right) ranked nodes with WLRank$_{1,2}$
for a random network of inertialess coupled oscillators with 1000 nodes 
  obtained by  rewiring  with probability $0.5$ a cyclic graph with constant nearest and next-to-nearest neighbor coupling 
   (supplementary materials, materials and methods). Each of the 12000 red crosses 
corresponds to one of 1000 random natural frequency vector $\bm P^{(0)}$ with components 
randomly distributed in $[-0.5,0.5]$ and 
summing to zero, multiplied by a prefactor $\beta = 0.4, 0.6, \ldots 2.6$. The blue crosses correspond to
running averages over 500 red crosses with consecutive values of ${\rm max}(\Delta \theta)$. 
}\label{fig:LR_WLR}
\end{figure}

In Fig.5 of the main text, we calculated the 
percentage of nodes with highest LRank$_2$ necessary to give the top 15 \% ranked nodes with WLRank$_2$.
The conclusions drawn from these data are generic -- they are valid for different percentages than 15\% and for
LRank$_1$ vs. WLRank$_1$. This is illustrated in 
Fig.~\ref{fig:LR_WLR}, which shows similar results for the percentage of nodes with highest
LRank$_{1,2}$ that include the top $10$\% and $20$\% ranked nodes with WLRank$_{1,2}$.

\pagebreak
\newpage

\end{document}